# Electron-phonon interactions and tensor analysis in topological insulator bismuth telluride using angle resolved polarized Raman spectroscopy

Aditya Singh[1], Anna Elsukova[2], Divya Rawat[1], Saswata Talukdar[3], Surajit Saha[3], Arnaud le Febvrier[4], Per O. Å. Persson[2], Per Eklund[4], and Ajay Soni[1,*]

[1]School of Physical Sciences, Indian Institute of Technology Mandi, Mandi 175005, Himachal Pradesh, India

[2]Thin Film Physics Division, Department of Physics, Chemistry, and Biology (IFM), Linköping University, Linköping, SE-58183, Sweden

[3]Department of Physics, Indian Institute of Science Education and Research, Bhopal 462066, India

[4]Department of Chemistry - Ångström Laboratory; Inorganic Chemistry, Uppsala University, Uppsala, 75105, Sweden

*Author to whom correspondence should be addressed: ajay@iitmandi.ac.in

**Abstract**

We report on the angle-resolved polarized Raman spectroscopy (ARPRS) and estimation of the Raman tensor elements using both classical and quantum treatments to analyse the polarized Raman spectra of single crystal $Bi_2Te_3$. The observed polar patterns and systematic variations in the relative intensities of four characteristic Raman-active modes indicate a higher differential polarizability along the c-axis, accompanied by anisotropic photon-phonon interactions. This interplay of electron-photon-phonon interactions is crucial for understanding the lattice dynamics of $Bi_2Te_3$, which underpin its thermoelectric performance and topological properties.

**Introduction:**

Bismuth telluride is a versatile material with significant applications in thermoelectricity, infrared photodetectors, energy storage, and photovoltaics [1-4]. It exhibits high thermoelectric figure of merit near room temperature arising from a combination of various factors, such as high band degeneracy, high effective mass and low lattice thermal conductivity. The discovery of topological surface states in $Bi_2Te_3$ [5] has further intensified the research interests due to its potential application in technologies like quantum computing and spintronics [6, 7]. Since the topological phenomena are intrinsically linked to electron-electron, electron-phonon (*e-ph*) [8], and exciton-photon interactions [9], a detailed understanding of the electron-phonon-photon interactions, phonon dynamics is crucial in $Bi_2Te_3$. The inherent structural anisotropy of $Bi_2Te_3$ allows tuning of its optical, electronic, and thermal properties through orientation engineering, thus unlocking new avenues for designing and developments of angle-resolved photonic and electronic devices. The impact of anisotropy on thermoelectricity, dielectric function, charge transport and sound speed has been extensively studied [10-13]. Raman spectroscopy investigations on $Bi_2Te_3$ have primarily focused on anharmonicity and its role in thermoelectric properties, interlayer vibrational modes, crystal symmetry breaking, and the evolution of optical phonons with temperature [14-17], however a comprehensive analysis of $Bi_2Te_3$ polarized Raman response using the Raman tensor formalism remains unexplored. Angle-resolved polarized Raman spectroscopy (ARPRS) offers a powerful tool to probe the role of *e-ph* interactions, extract the Raman tensor elements, and elucidate the fundamental light matter interactions across different crystal orientations. This approach provides critical insights into the anisotropic nature of light-matter interactions in $Bi_2Te_3$, paving the way for its applicability in advanced quantum devices.

Photon-phonon interactions driven by structural anisotropy have been extensively studied in two-dimensional materials, particularly transition metal dichalcogenides, using polarised Raman tensor formalism. Although this Raman tensor formalism has been applied to materials like black phosphorus, $MoS_2$, $MoSe_2$, $Bi_2Se_3$, and $ReSe_2$ [18-22], a comprehensive analysis of polarized Raman response of $Bi_2Te_3$ remains lacking. This methodology is a powerful tool for probing multibody interactions, including electron-electron, (*e-ph*), exciton, plasmon, and phonon-plasmon coupling. These fundamental interactions underpin the operation of a wide range of advanced technologies, such as photonic crystals, optoelectronic devices, photovoltaics, and plasmonic devices [23, 24]. Especially, the anisotropy in light-matter interactions sheds light on differential polarizability and the role of of *e-ph* interactions in these advanced materials. In this study, we investigated the influence of lattice-driven anisotropy on light-matter interactions by estimating the Raman tensor elements and role of *e-ph* interactions through a quantum treatment of Raman tensor analysis. A correlation between lattice vibrations and associated differential polarizability of the phonon modes has been established for $Bi_2Te_3$. This approach to Raman tensor analysis can be further extended to the broader family of topological quantum materials, enabling a deeper understanding of inelastic light scattering processes and the anisotropic electron-photon-phonon interactions.

**Experimental Section:**

Single crystal of $Bi_2Te_3$ was synthesized with stoichiometric amounts of bismuth and tellurium (make: Alpha Aesar, 5N purity) placed in a quartz tube and vacuum sealed to a pressure of $< 10^{-5}$ mbar. A dual zone Bridgman furnace was used with a ~100° C thermal gradient between the hot zone (~ 650° C) and cold zone (~ 550° C). The quartz tube was pulled with a speed of ~ 2 mm/hr for ~ 14 days from hot zone to cold zone. In order to determine the phase purity, X-Ray diffraction (XRD) was carried out using Rigaku Smartlab X-ray diffractometer with $CuK_\alpha$ radiation (wavelength ~1.5406 Å) in Bragg-Brentano geometry[22]. Rietveld refinement was carried out using Fullprof software in order to determine phase purity as well as the lattice parameters[25]. To study the morphology, field emission scanning electron microscope (FESEM, JFEI, USA, Nova Nano SEM 450) was used. The transmission electron microscopy (TEM) including high-resolution scanning TEM (HR-STEM) imaging was performed in the Linköping double-corrected FEI Titan$^3$ 60–300, operated at 300 kV. A cross-sectional TEM lamella was prepared, using focused an ion beam instrument (Carl Zeiss Cross-Beam 1540 EsB system). The diffraction pattern was analysed and indexed by matching with simulations obtained using SINGLE CRYSTAL$^{TM}$ software by Crystal Maker Software Ltd, Oxford, England. The crystal structure was visualized using VESTA software. Room temperature Raman spectroscopy measurements were carried out using a Horiba LabRAM HR Evolution Raman spectrometer in backscattering geometry. A solid-state laser excitation source of 532 nm with Czerny-turner grating (1800 grooves/mm) and a Peltier cooled CCD detector were used for acquisition of Raman data. Ultra-low frequency filters were used to access the low frequency Raman modes close to the excitation laser (~ 20 cm$^{-1}$). As $Bi_2Te_3$ is a poor thermal conductor, so the laser power was kept very low (~20 $\mu W$) to avoid local heating of the sample, leading to a larger error bar in the measurements. The polarized Raman measurements were carried out by placing a half wave plate in the path of incident light and an analyser before the detector. The sample was placed on a 360° rotating stage and rotated in step of 10°.

**Results and discussion:**

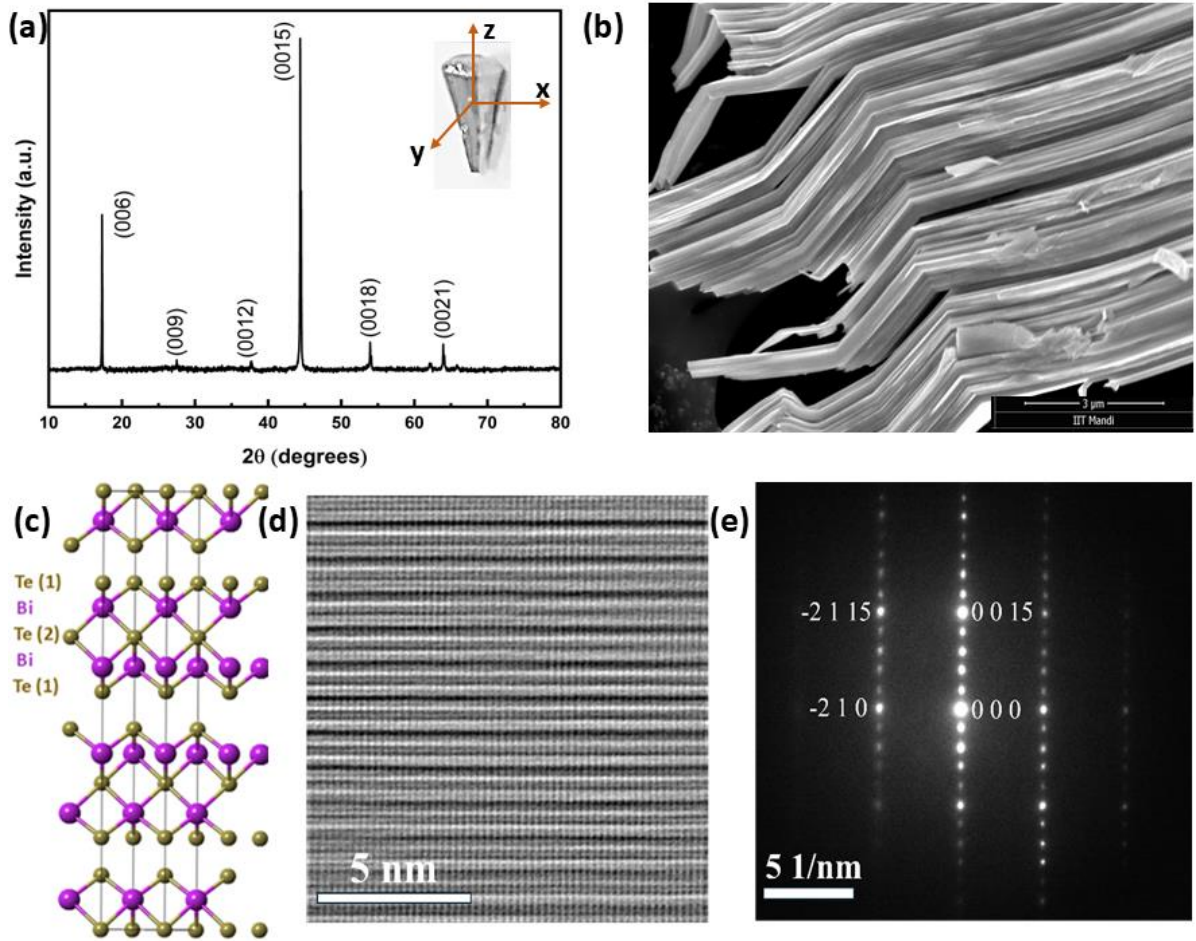

Fig 1. (a) XRD of single crystal $Bi_2Te_3$, inset showing the optical photograph, (b) SEM image of fractured surface, (c) Crystal structure viewed in [120] direction, (d) HAAD-STEM image and (e) SAED pattern viewed along [1 2 0] zone axis.

The XRD (Fig 1(a)) reveals the crystalline nature as well as the unidirectional orientation of the crystal along the c axis, with primarily reflections along (0 0 3n) directions. The lattice parameters obtained from the Rietveld refinement are, a = b ~ 4.38 Å and c ~ 30.50 Å with a unit cell volume ~ 508.10 Å$^3$, which are matching with the JCPDS Card No. 82-358 (see Fig S1 in supplemental material file)[26] [27]. The FESEM image (Fig 1 (b)) of the fractured surface shows the layered nature of the crystal. $Bi_2Te_3$ crystallizes in a layered crystal structure in rhombohedral geometry with repeating Bi and Te atomic layers, as shown in Fig. 1(c). These atomic layers are comprised of quintuple arrangement of atoms in Te(1) – Bi – Te(2) – Bi – Te(1), where 1 and 2 represents the different chemical environment around Te atoms. In $Bi_2Te_3$, successive quintuple layers are weakly held together by van der Waals forces, while the two atomic layers within each quintuple layer are strongly bonded through a combination of covalent and ionic interactions. The Bi-Te(1) bond exhibits a slightly ionic character, whereas the Bi-Te(2) bond is predominantly covalent[28]. Atomically resolved High Angular Annular Dark Field (HAADF) STEM image (Fig.1 (d)) shows the highly ordered structure with gaps separating quintuple layers along the *c*-axis. The atomic intensity variations are also consistent with the expected atomic arrangements (Fig 1 (c)). Presence of expected structure was further confirmed by Selected Area Electron Diffraction (SAED) (Fig. 1 (e)). The lattice parameters obtained from the SAED were found to be c = 32.22 ± 0.07 Å, a = b = 4.65 ± 0.01 Å are consistent with the ones obtained from Rietveld refinement.

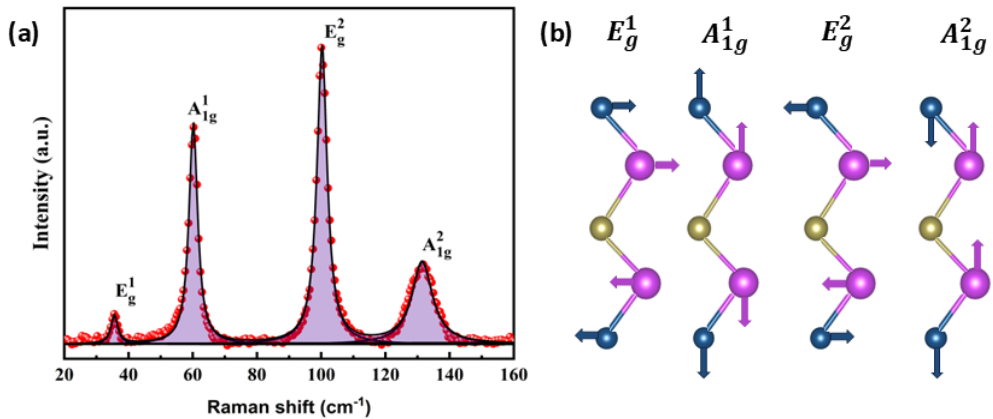

Fig 2: (a) Room temperature Raman spectra of $Bi_2Te_3$ and (d) Schematic of atomic displacements of $E_g$ and $A_{1g}$ modes.

Room temperature Raman spectra (Fig 2(a)) has four characteristic Raman modes present at ~ 37 $cm^{-1}(E_g^1)$, ~ 62 $cm^{-1}(A_{1g}^1)$, ~ 102 $cm^{-1}(E_g^2)$, ~ 133 $cm^{-1}(A_{1g}^2)$ [28]. There are five atoms in one primitive unit cell of $Bi_2Te_3$ which correspond to 15 zone-centre vibrational modes, out of which 12 are optical and 3 are acoustic modes. The irreducible representation of these zone centre phonon modes is given by $\chi = 2E_g + 2A_{1g} + 2E_u + 2A_{1u}$ where g and u represent Raman active and infrared active modes respectively. According to the group theory analysis, [29] these optically active modes are characterized into $E_g$ and $A_{1g}$ symmetries, which correspond to in-plane and out-of-plane vibration, respectively (Fig 2(d)). Here, the $A_{1g}^1$ and $E_g^1$ modes represent symmetric atomic displacements while the $A_{1g}^2$ and $E_g^2$ modes represent antisymmetric stretching vibrations. In general, any Raman active mode (say $\mu^{th}$) is associated with a specific Raman tensor($R_{ij}^{\mu} = |R_{ij}^{\mu}|e^{i\phi_{ij}^{\mu}}$ ), which is a 3×3 matrix comprising of Raman tensor elements($|R_{ij}^{\mu}|$) and associated phase($\phi_{ij}^{\mu}$) [30]. For $Bi_2Te_3$, the Raman tensor of $\mu^{th}$ mode is given by

$$R(A_{1g}) = \begin{bmatrix} \eta e^{i\phi_\eta} & 0 & 0 \\ 0 & \eta e^{i\phi_\eta} & 0 \\ 0 & 0 & \beta e^{i\phi_\eta} \end{bmatrix}$$

$$R(E_g) = \begin{bmatrix} \gamma e^{i\phi_\gamma} & 0 & 0 \\ 0 & -\gamma e^{i\phi_\gamma} & \delta e^{i\phi_\delta} \\ 0 & \delta e^{i\phi_\delta} & 0 \end{bmatrix}$$

$$R(E_g) = \begin{bmatrix} 0 & -\gamma e^{i\phi_\gamma} & \delta e^{i\phi_\delta} \\ -\gamma e^{i\phi_\gamma} & 0 & 0 \\ \delta e^{i\phi_\delta} & 0 & 0 \end{bmatrix}$$

Here $\eta, \beta, \gamma, \delta$ corresponds to Raman tensor elements and $\phi_\eta, \phi_\beta, \phi_\gamma, \phi_\delta$ denotes the complex phases of Raman tensor elements [29]. As $Bi_2Te_3$ is an anisotropic material with high anisotropy ratio (c/a ~ 6.96), the quantum interactions are expected to be different with crystallographic orientation, which can be analysed by opting different experimental configuration such as incident light falling on either basal plane or *m*-plane orientations. Hence ARPRS is an important tool to study these multibody interactions with respect to the crystal orientation, thereby quantifying the anisotropic nature of light matter interactions present in $Bi_2Te_3$.

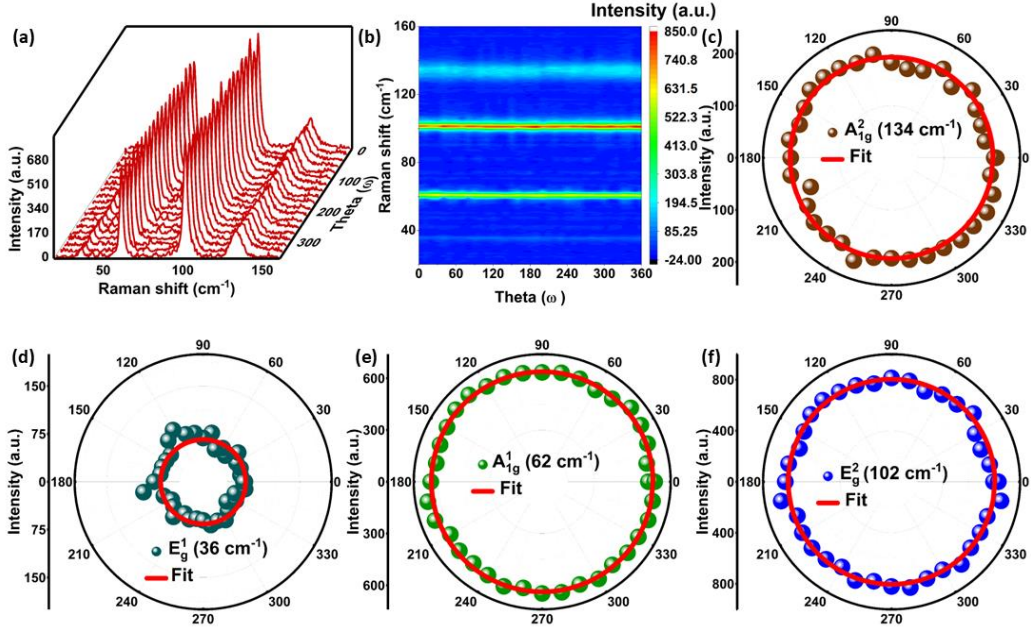

Fig 3. (a) ARPRS for basal plane configuration, (b) corresponding colour plot of intensity and (c-e) Polar plots of $A_{1g}^2$, $E_g^1$, $A_{1g}^1$, $E_g^2$, modes, respectively. The solid circles are the experimental data points and the red line denotes the fitting of the polar plots using Placzek Intensity approximation.

Fig 3(a) shows the room-temperature polarized Raman spectra and the corresponding intensity contrast plot are presented in Fig 3(b). These spectra are obtained by rotating the sample on a stage from 0° to 360°, with 10° step size, in the basal plane. The experimental configurations of ARPRS with the orientation of the sample are shown in the supplemental material file Fig S2 [26]. When the polarized laser incidents on the *ab* plane (parallel to c axis), it is referred as the basal plane configuration whereas when the light is falling on *ac*/*bc* plane, we are referring it as *m*-plane (perpendicular to c axis). The intensity of all the modes in polar plots (Fig 3(c-f)), remains constant with the angle of rotation across the basal plane.

Table I: Intensity of Raman modes using Placzek intensity approximation equation

| S.No. | Configuration | Raman scattering intensity |
|---|---|---|
| 1. | Basal plane | $I_{A_{1g}} = \|\eta\|^2$ |
|  |  | $I_{E_g} = \|\gamma\|^2$ |
| 2. | *m*- Plane | $I_{A_{1g}} = \|\eta\|^2 sin^4\theta + \|\beta\|^2 cos^4\theta + \frac{1}{2}\|\eta\|\|\beta\|sin^2(2\theta)cos_{\varphi_{\eta\beta}}$ |
|  |  | $I_{E_g} = \|\gamma\|^2 cos^4\theta + \|\delta\|^2 sin^2 2\theta - \|\gamma\|\|\delta\| sin(2\theta) cos^2\theta \times 2cos_{\varphi_{\gamma\delta}}$ |

To estimate the Raman tensor elements and analyse the intensity *(I)* variation with the angle of polarized light, the obtained data have been fitted with the Placzek intensity approximation [31], which can be written as $I = |e_i.R.e_s|^2$. Here $e_i$ and $e_s$ denotes the polarization vectors of incident and scattered light and *R* represents the characteristic Raman tensor of the modes [32]. When the incident light is falling on the basal plane, these vectors take the form of $e_i$ = (sin ω, cos ω, 0) and when the light is falling on the *m*-plane the polarization vectors take the form of $e_s$= (0, sin θ, cos θ) where ω and θ represents the angle which the incident light makes with the basal/ *m*-plane respectively. On substituting the respective Raman tensor elements of the $\mu^{th}$ mode in the Placzcek equation, we get the mathematical equations as tabulated in Table I. By analysing the experimental data with mathematical formulations, we estimated the anisotropy ratio, relative differential polarizability, and *e-ph* interactions, [22] based on the mode symmetry and their coupling with the incident light.

For the basal plane configuration, the intensity depends on the square of the Raman tensor elements and remain constant irrespective of the sample rotation angle as well as associated phase term, which indicates to isotropic light-matter interactions (Fig 3(c-f)). Since the propagation vector of the incident light aligns with the c-axis, there is no polarization component along the c-axis. Consequently, the β parameter of the out-of-plane $A_{1g}$ mode cannot be evaluated while it is zero for $E_g$ modes. From the fitting of the isotropic intensity in basal plane, we estimated the tensor elements $E_g$ (~ 8 ($E_g^1$) and ~ 28 ($E_g^2$)) and $A_{1g}$ (~ 25 ($A_{1g}^1$) and ~ 13 ($A_{1g}^2$)), with zero phase factor ($\phi_\eta$ = 0) and the details have been presented in Table II.

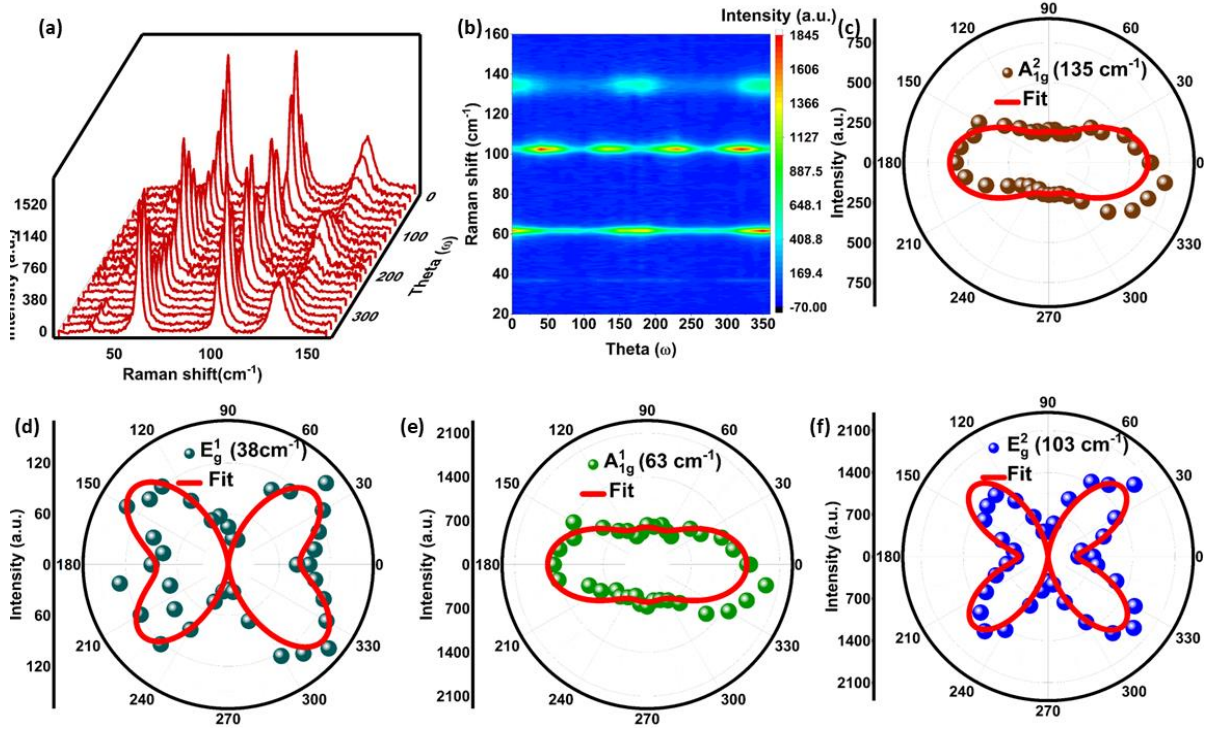

Fig 4. (a) ARPRS for *m*-plane configuration, (b) corresponding colour plot of intensity, (c-e) Polar Plots of $A_{1g}^2$, $E_g^1$, $A_{1g}^1$, $E_g^2$, modes, respectively. The red line denotes the fitting of the polar plots using Placzek Intensity approximation.

Table II: Raman tensor elements estimated using Placzek Approximation.[31]

| | In Plane | | Cross Plane |
|---|---|---|---|
| $E_g^1$ | $\begin{pmatrix} 8 & -8 & \delta \\ -8 & -8 & \delta \\ \delta & \delta & 0 \end{pmatrix}$ | $E_g^1$ | $\begin{pmatrix} 9 & -9 & -10e^{i0.26\pi} \\ -9 & -9 & 10e^{i0.26\pi} \\ -10e^{i0.26\pi} & 10e^{i0.26\pi} & 0 \end{pmatrix}$ |
| $A_{1g}^1$ | $\begin{pmatrix} 25 & 0 & 0 \\ 0 & 25 & 0 \\ 0 & 0 & \beta \end{pmatrix}$ | $A_{1g}^1$ | $\begin{pmatrix} 24 & 0 & 0 \\ 0 & 24 & 0 \\ 0 & 0 & 39e^{i0.68\pi} \end{pmatrix}$ |
| $E_g^2$ | $\begin{pmatrix} 28 & -28 & \delta \\ -28 & -28 & \delta \\ \delta & \delta & 0 \end{pmatrix}$ | $E_g^2$ | $\begin{pmatrix} 21 & -21 & -39e^{i0.55\pi} \\ -21 & -21 & 39e^{i0.55\pi} \\ -39e^{i0.55\pi} & 39e^{i0.55\pi} & 0 \end{pmatrix}$ |
| $A_{1g}^2$ | $\begin{pmatrix} 13 & 0 & 0 \\ 0 & 13 & 0 \\ 0 & 13 & \beta \end{pmatrix}$ | $A_{1g}^2$ | $\begin{pmatrix} 13 & 0 & 0 \\ 0 & 13 & 0 \\ 0 & 0 & 24e^{i0.18\pi} \end{pmatrix}$ |

Fig 4 (a-b) illustrates the Raman spectra and the associated intensity colour plot when the incident light is falling on the *m*-plane. A distinct periodic variation in intensity of the Raman modes (Fig 4(c-f)) has been observed with respect to the angular rotation. The contrasting behaviour of the $E_g$ and $A_{1g}$ modes indicates the anisotropic interactions of light with $Bi_2Te_3$ crystals in *m*-plane. The similar pattern has been observed for ~ 633 nm excitation, (Fig S3, Supplemental material file)[26], confirming the intrinsic nature of structurally driven (an/)isotropic light-matter interactions in these materials. The mode frequency as well as full width half maxima for 532 nm, with respect to the angle of polarized light shows negligible variation as shown in Fig S4 (Supplemental material )[26].

The tensor elements extracted from the fitting of the intensity for the *m*-plane are $E_g$ (~ 9 & ~10 ($E_g^1$) and ~ 21 & ~ 39 ($E_g^2$)) and $A_{1g}$ (~ 24 & ~ 39 ($A_{1g}^1$) and ~ 13 & ~24 ($A_{1g}^2$)). The phase associated with the *m*-plane elements are ~ 0.26 π ($E_g^1$), ~ 0.68 π ($A_{1g}^1$), ~ 0.55 π ($A_{1g}^2$), ~ 0.18 π ($E_g^2$), respectively. The ratio $\left(\frac{\beta}{\eta} \sim 1.625\right)$ for $A_{1g}^1$ mode denotes the presence of high anisotropy in $Bi_2Te_3$[21]. Here the nature of anisotropic light matter interactions is related to the differential polarizability, which can be given by the equation,

$$R_{ij}^q = V_{prim} \sum_{p=1}^{N} \sum_{l=1}^{3} \frac{\partial \alpha_{ij}}{\partial r_l(p)} \frac{e_l^q(p)}{\varepsilon_0 \sqrt{M_p}}.$$

Here $R_{ij}^q$ is the characteristic Raman tensor of phonon mode *q*, $r_l(p)$ is the *l* coordinate of the $p^{th}$ atom, $e_l^q(p)$ is the displacement of atom $p^{th}$ in the direction *l*, $V_{prim}$ is the volume of primitive unit cell of the material, $M_p$ is the mass of the atom and $\alpha_{ij}$ is the polarizability tensor[32]. Thus, the inelastic light scattering for the ($A_{1g}$) mode suggests higher differential polarizability along the *c* axis as the Raman tensor elements are greater ($\beta > \eta$, 39 > 24 for $A_{1g}^1$ and 24 >13 for $A_{1g}^2$). Furthermore, the ratio of the Raman tensor elements of the $E_g^1$ modes and the $A_{1g}^1$ modes ($\frac{\gamma}{\eta}$) remains less than unity, which signifies the differential polarizability of the $A_{1g}^1$ mode is greater than that of the $E_g^1$ modes[18]. Therefore, the $A_{1g}$ mode with out of plane vibrations has strong sensitivity towards polarized light and distinct Raman signature [33].To ensure the reliability of the observed results, we calculated the ratio of Raman tensor elements for the $E_g^1$ ($\frac{\gamma_{m-plane}}{\gamma_{basal-plane}}$) and $A_{1g}^1$ ($\frac{\eta_{m-plane}}{\eta_{basal-plane}}$) modes in both cross-plane and in-plane configurations. These ratios were found to be approximately 1.125 and 0.96 respectively. The minimal variation in these values across different experimental configurations (basal and

*m*-plane) indicates that the Raman tensor elements remain consistent. The estimated Raman tensor elements obtained from intensity fitting in Fig 4(c-f) are shown in Table II. Minor variations in the polar plot of $A_{1g}$ mode at 90° and 270° indicate a contribution from second-order susceptibility (macroscopic description) or *e-ph* interactions (microscopic description) to the Raman scattering[22].

Macroscopically, the bulk polarization (P) induced by incident light ($E^i$) is given by $P = \epsilon_0 \overleftrightarrow{\chi}.E^i$, where $\epsilon_0$ is the electric permittivity of free space and $\overleftrightarrow{\chi}$ is the second-rank tensor of electronic susceptibility [30]. On the other hand, a quantum mechanical treatment of the Raman tensor is required to address the microscopic processes by incorporating energy exchange between the incident and scattered light, along with electron-photon and electron-phonon interactions.[31]. The quantum formalism of the Raman tensor elements of the μ$^{th}$ mode can be expressed as: -

$$R_{ij}^\mu = \frac{1}{V} \sum_{v,c,c'} \sum_{\vec{q}} \frac{\langle \psi_v(\vec{q})|\vec{e}_s.\vec{\nabla}|\psi_{c'}(\vec{q})\rangle \langle \psi_{c'}(\vec{q})|H_{ep}^k|\psi_c(\vec{q})\rangle \langle \psi_c(\vec{q})|\vec{e}_i.\vec{\nabla}|\psi_v(\vec{q})\rangle}{(E_L - E_{cv}(\vec{q}) - i\Gamma_c)(E_L - \hbar\omega_{ph}^\mu - E_{c'v}(\vec{q}) - i\Gamma_{c'})}$$

where V is the crystal volume. The summation is carried over all branches of conduction and valence bands within the first Brillouin zone. The numerator comprises of two electron photon matrix elements, $\langle \psi_v(\vec{q})|\vec{e}_s.\vec{\nabla}|\psi_{c'}(\vec{q})\rangle$ for scattered, and $\langle \psi_c(\vec{q})|\vec{e}_i.\vec{\nabla}|\psi_v(\vec{q})\rangle$ for incident in addition to one *e-ph* matrix element $\langle \psi_{c'}(\vec{q})|H_{ep}^k|\psi_c(\vec{q})\rangle$. Here $\Gamma_c$ and $\Gamma_{c'}$ are representing the broadening factor connected with the lifetime of photoexcited states[31]. The variation in the polar plots of $A_{1g}$ mode is attributed to the anisotropic nature of electron-phonon interactions[19, 20, 22, 34]. The complex nature of the Raman tensor can be elucidated through a detailed analysis of macroscopic and microscopic Raman scattering processes [30]. The combined macroscopic and microscopic approach provides a fundamental understanding of the anisotropic light-matter interactions in $Bi_2Te_3$ due to the second-order susceptibility or *e-ph* interactions.

Thus, the basal plane and *m*-plane configuration of ARPRS reveal the presence of isotropic and anisotropic light matter interactions, akin to the studies on $MoSe_2$, $MoS_2$, $Bi_2Se_3$ and $WSe_2$ [18, 21, 22, 35]. The investigation advances the fundamental understanding of microscopic scattering processes and their directional dependence in $Bi_2Te_3$ for topological insulators as well as thermoelectricity.

**Conclusion:**

In summary we have estimated the fundamental Raman tensor elements to understand the role of anisotropic differential polarizability and fundamental multibody phenomenon including *e-ph* interactions in $Bi_2Te_3$. A correlation between lattice dynamics and light matter interactions has been established for $Bi_2Te_3$. The Raman tensor analysis can be further extended to the other quantum materials for a better understanding of inelastic light scattering processes and the anisotropic electron-photon-phonon interactions.

**ACKNOWLEDGMENTS:** A.S and A.S. thank IIT Mandi for instruments and research facilities and acknowledge DST India for Indo-Sweden bilateral grant (Grant No. DST/INT/SWD/VR/P-18/2019). The funding is acknowledged from the Swedish Government Strategic Research Area in Materials Science on Functional Materials at Linköping University (Faculty Grant SFO-Mat-LiU No. 2009 00971), the Knut and Alice Wallenberg foundation through the Wallenberg Academy Fellows program (KAW-2020.0196, P.E.), the Swedish Research Council (VR) under Project No. 2021-03826 (P.E.). A.E. and P.O.Å.P would like to acknowledge access to the Swedish National Infrastructure for Advanced Electron Microscopy, ARTEMI, supported by the Swedish Research Council (VR) and the Foundation for Strategic Research (SSF), through Grants No. 2021-00171 and No. RIF21-0026.

30. J. Kim, J.-U. Lee, and H. Cheong, *Polarized Raman spectroscopy for studying two-dimensional materials.* Journal of Physics: Condensed Matter, 2020. **32**(34): p. 343001.
31. R. Loudon, *The Raman effect in crystals.* Adv. Phys., 1964. **13**(52): p. 423-482.
32. M. Cardona and G. Güntherodt, *Light Scattering in Solids II: Basic Concepts and Instrumentation*. 1982: Springer-Verlag.
33. J. Zhang, Z. Peng, A. Soni, Y. Zhao, Y. Xiong, B. Peng, J. Wang, M. S. Dresselhaus, and Q. Xiong, *Raman Spectroscopy of Few-Quintuple Layer Topological Insulator $Bi_2Se_3$ Nanoplatelets.* Nano Letters, 2011. **11**(6): p. 2407-2414.
34. S. Zhang, N. Mao, N. Zhang, J. Wu, L. Tong, and J. Zhang, *Anomalous Polarized Raman Scattering and Large Circular Intensity Differential in Layered Triclinic $ReS_2$.* ACS Nano, 2017. **11**(10): p. 10366-10372.
35. M. Jin, W. Zheng, Y. Ding, Y. Zhu, W. Wang, and F. Huang, *Raman Tensor of $WSe_2$ via Angle-Resolved Polarized Raman Spectroscopy.* The Journal of Physical Chemistry C, 2019. **123**(48): p. 29337-29342.